\documentclass[twocolumn,aps,prl,groupedaddress]{revtex4}
\usepackage{epsfig,amssymb,amsmath}
\usepackage[hypertex,linkcolor=red]{hyperref}
\def\comment#1{}\def\labell#1{\label{#1}}
\begin{document}
\title{Quantum private queries}
  
\author{Vittorio Giovannetti$^1$, Seth Lloyd$^2$, and Lorenzo
  Maccone$^3$}\affiliation{$^1$NEST-CNR-INFM \& Scuola Normale
  Superiore, Piazza dei Cavalieri 7, I-56126, Pisa, Italy.\\$^{2}$MIT,
  RLE and Dept. of Mech. Engin. MIT 3-160, 77 Mass.~Av., Cambridge, MA
  02139, USA.\\ $^3$QUIT, Dip.  Fisica ``A.  Volta'', Univ. Pavia, via
  Bassi 6, I-27100 Pavia, Italy.}
\begin{abstract}
  We propose a cheat sensitive quantum protocol to perform a private
  search on a classical database which is efficient in terms of
  communication complexity.  It allows a user to retrieve an item from
  the server in possession of the database without revealing which
  item she retrieved: if the server tries to obtain information on the
  query, the person querying the database can find it out.
  Furthermore our protocol ensures perfect data privacy of the
  database, i.e. the information that the user can retrieve in a
  single queries is bounded and does not depend on the size of the
  database.  With respect to the known (quantum and classical)
  strategies for private information retrieval, our protocol displays
  an exponential reduction both in communication complexity and in
  running-time computational complexity.
\end{abstract}
\maketitle 


Privacy is a major concern in many information transactions.  A
familiar example is provided by the transactions between web search
engines and their users.  On one hand, the user (say Alice) would
typically prefer not to reveal to the server the item she is
interested in ({\em user privacy}). On the other hand, the server (say
Bob) would like not to disclose more information than that Alice has
asked for ({\em data privacy}).  User and data privacy are apparently
in conflict: the most straightforward way to obtain user privacy is
for Alice to have Bob send her the entire database, leading to no data
privacy whatsoever.  Conversely, techniques for guaranteeing the
server's data privacy typically leave the user vulnerable~\cite{SPIR}.
At the information theoretical level, this problem has been formalized
by Gertner {\em et al.}  as the Symmetrically-Private Information
Retrieval (SPIR)~\cite{SPIR}.  This is a generalization of the Private
Information Retrieval (PIR) problem~\cite{PIR} which deals with user
privacy alone.  (SPIR is closely related to oblivious
transfer~\cite{oblivious}, in which Bob sends to Alice $N$ bits, out
of which Alice can access exactly one--which one, Bob doesn't know.)
No efficient solutions in terms of communication
complexity~\cite{amba} are known for SPIR.  Indeed, even rephrasing
them at a quantum level~\cite{kerend1,kerend2}, the best known
solution for the SPIR problem (with a single database server) employs
$O(N)$ qubits to be exchanged between the server and the
user~\cite{NOTA} and ensures data privacy only in the case of honest
users (here $N$ is the number of items contained in the database,
while an honest user is defined as one who does not want to compromise
her chances of getting the information about the selected item in
order to get more).  PIR admits protocols that are more efficient in
terms of communication complexity~\cite{PIR}.  \comment{In reality,
  there doesn't seem to be any single server PIR or SPIR that has
  communication complexity less than $O(N)$!}  As will be seen below,
however, both PIR and SPIR necessarily require $O(N)$ computational
complexity on the part of the database.

In this paper we present a new quantum cryptographic
primitive~\cite{review1}, the quantum private query (QPQ), which
allows an exponential reduction in the communication and computational
complexity with respect to the best (quantum or classical) SPIR
protocol proposed so far.  QPQ ensures perfect data privacy and it
exploits a cheat sensitive strategy~\cite{hardy} that allows Alice to
determine whether Bob has been trying to cheat to obtain information
about her query.  In other words, Alice can ask Bob's database a
question and obtain the answer, together with a quantum certificate
that Bob retains no record of what question she asked.  With respect
to (classical or quantum) SPIR and oblivious transfer protocols QPQ
presents an exponential reduction in communication complexity. This
comes from the fact that information theoretic SPIR protocols require
the exchange of the whole database~\cite{NOTA}, $O(N)$ qubits, while
QPQ requires the exchange of only two database elements, identified by
$O(\log N)$ qubits.  Quantum Private Queries also provides an
exponential reduction in computational complexity over all classical
PIR schemes, whether symmetric or not.  In both cryptographic and
information-theoretic PIR protocols, the owner(s) of the database(s)
must perform $O(N)$ `internal' database calls in response to Alice's
query.  That is, as part of the protocol, Bob must perform operations
that access {\it every} entry in his database, using some
cryptographic primitive such as a public key supplied by Alice.  If
the PIR protocol requires Bob to perform fewer than $N$ internal
database calls, then he obtains information about Alice's query simply
by monitoring which database entries were and were not called in the
course of executing the protocol.  That is, a classical PIR protocol
necessarily has database computational complexity $O(N)$ per query.
In contrast, Quantum Private Queries require only two internal
database calls per use, each using only $O(\log N)$ time
steps~\cite{qram}.

Quantum private queries achieve two competing goals: Bob can provide
the service of private searching without having to give up his
database, and Alice can test his honesty without having to trust him.
The basic idea underlying the protocol is simple: Bob, as a sign of
his discretion, returns not only the answer to Alice's query, but the
original query itself, retaining no copy.  Alice, in addition to
performing normal queries, can perform also quantum superpositions of
different queries. This means that in addition to being able to
request the $j$th or the $k$th records in the database, she can also
request both records in a quantum superposition.  To find out whether
Bob is trying to discover her queries, she just has to send proper
superpositions of queries and check Bob's answer to see whether the
superposition has been preserved. In this case, she can be confident
that Bob has retained no information about her query: any capture of
information by Bob would have induced a disturbance. The user security
rests on Bob's impossibility of discovering the generic quantum state
of Alice's query.  Two basic elements of quantum theory enforce this:
the no-cloning theorem~\cite{noclon} which forbids the discovery of
the state starting from a single copy of it~\cite{mauro}, and the
inability fully to characterize a composite system using only local
operations.  The database security of QPQ is ensured by the finite
number of signals Bob is sending back to Alice. As we will see these
can be as low as two. This automatically implies that in the QPQ a
dishonest Alice will be able to recover at most two items from the
database to be compared with the $O(\log N)$ bits of information a
dishonest user will be able to acquire in the quantum SPIR
protocols~\cite{kerend2}.

The rest of this paper is devoted to making the previous ideas rigorous
and to providing the details of the  protocols. We start by
describing the quantum communication protocol that Alice and Bob must
follow, and give a security analysis. We then conclude
with a discussion on how Bob can
interrogate his database preserving Alice's superposed queries.

\begin{figure}[t!]
\begin{center}
\epsfxsize=.7\hsize\leavevmode\epsffile{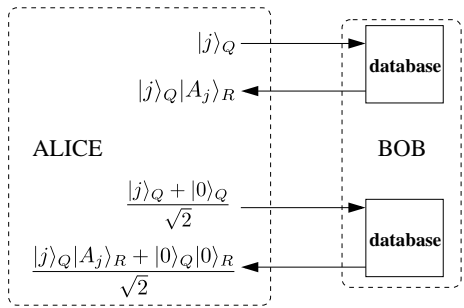}
\end{center}
\vspace{-.5cm}
\caption{Scheme of the QPQ protocol. Alice wants to find
  out the $j$th record of Bob's database. She then prepares two
  $n$-qubit registers, one contains the state $|j\rangle_Q$, the other
  contains the quantum superposition
  $(|j\rangle_Q+|0\rangle_Q)/\sqrt{2}$. (She knows that the $0$th
  record of Bob's database contains the fixed value $A_0=0$). She then
  sends, in random order, these two registers to Bob, waiting for his
  first reply before sending the second. Bob uses each of the two
  registers to interrogate his database using a qRAM device, which
  records the reply to her queries in a register $R$. At the end of
  their exchange, Alice possesses the states
  $|j\rangle_Q|A_j\rangle_R$ and
  $(|j\rangle_Q|A_j\rangle_R+|0\rangle_Q|0\rangle_R)/\sqrt{2}$, where
  the $A_j$ is the content of the $j$th record in the database.  By
  measuring the first she obtains the value of $A_j$, with which she
  can check whether the superposition in the second state was
  preserved. In this case she can be confident Bob obtained no
  information on what $j$ was. }  \labell{f:proto}
\end{figure}

To submit her query on the $j$th record of Bob's database, Alice uses
an $n$ qubit memory register $Q$.  It allows her to interrogate a
database of up to $N=2^n$ elements. To test whether Bob is cheating
and is trying to find out what her query is, she needs to submit a
superposition of queries. So she prepares two copies of the register
$Q$, one is initialized as $|j\rangle_Q$, the other as
$(|j\rangle_Q+|0\rangle_Q)/\sqrt{2}$ (we suppose that the $0$th record
in Bob's database contains a fixed reference value known to her). She
then {\em randomly} chooses one of these two registers and sends it to
Bob. He interrogates his database using it as an index register
employing the qRAM algorithm described below [see Eq.~(\ref{inout})].
It returns a second register $R$ which contains the answer to the
query, and which may be entangled with the register $Q$ if the latter
was in the superposition state (without loss of generality we can
assume $R$ to be a single qubit). Bob sends back the $Q$ and $R$
registers to Alice. She then sends him her second $Q$ register, which,
again, is employed by Bob to interrogate his database and sent back to
Alice together with a new $R$ register containing the answer to her
second query. It is important to stress that Bob never knows if the
register he receives from Alice is the one containing the quantum
superposition or the other one: this means he does not know which
measurement could extract information on $j$ without disturbing the
register.  The number of exchanged qubits is $2 (n +1)= 2 (\log N +1)
$ (of these only $2$ contain information on the database).
We see that, in attempting to obtain information about Alice's state, Bob must
try to distinguish between two possible states that have overlap
$1/\sqrt 2$.  That is, Bob's position is isomorphic to that of
Eve in conventional quantum cryptography, and any attempt on
his part to gain information must necessarily be detected by
Alice: the tradeoff between the information that Bob can obtain    
and his probability of being detected by Alice are essentially
the same as in quantum cryptography (see, e.g.,~\cite{CWINTER})
as we now demonstrate.

After this double exchange with Bob, Alice is in
possession of the two states $|\psi_1\rangle=|j\rangle_Q|A_j\rangle_R$
and
\begin{eqnarray}
|\psi_2\rangle=\frac 1{\sqrt{2}}\Big(|j\rangle_Q|A_j\rangle_R+
|0\rangle_Q|A_0\rangle_R\Big)\labell{st}\;,
\end{eqnarray}
where $A_m$ is the content of the $m$th record in the database
(without loss of generality we can suppose that $A_0=0$). She can
recover the value of $A_j$ by measuring $|\psi_1\rangle$. This value
answers her query, and can be used to construct a measurement to test
whether the second state is really of the form $|\psi_2\rangle$ given
in Eq.~(\ref{st}). We will show that if Bob is acquiring information
on $j$, he will be perturbing the superposition state $|\psi_2\rangle$
and Alice has a nonzero probability of finding it out.  
The only
assumption necessary (which may be dropped by complicating the
protocol slightly) is that the value $A_j$ is uniquely determined by
$j$, i.e.~that there cannot be two different answers to one query.

The simple protocol described here can be easily modified to increase
its performance. First of all, in place of the fixed superposition
$(|j\rangle_Q + |0\rangle_Q)/\sqrt{2}$, we can allow Alice to employ
any arbitrary superposition $\alpha |j\rangle + \beta |0\rangle$ with
complex amplitudes $\alpha$ and $\beta$ unknown to Bob.  In this way
Bob's ability of masking his actions is greatly reduced.  More
generally, instead of creating a superposition with the reference
query $|0\rangle_Q$, she could superimpose two (or more) different
queries.  In this case, in addition to the query $j$ which she is
interested in, she randomly chooses another query (say the $k$-th).
Now she prepares three $n$-qubits registers in the state $|j\rangle$,
$|k\rangle$, and $(|j\rangle+|k\rangle)/\sqrt{2}$.  As in the case
discussed previously, she sends the registers to Bob in random order
and one-by-one (i.e.~she waits for Bob's reply before submitting the
next). At the end of their exchange, if Bob has not cheated, Alice is
in possession of three states: i.e.~$|j\rangle|A_j\rangle$, $|k\rangle
|A_{k}\rangle$, and
$(|j\rangle|A_j\rangle+|k\rangle|A_{k}\rangle)/\sqrt{2}$.  She starts
by measuring the first two, in order to find out the values of $A_j$
and $A_{k}$: the former is the answer she was looking for, the latter
will be used to prepare a measurement to test the third state to see
whether the superposition has been retained. In this case she can
conclude that Bob has not cheated.  Notice that, in contrast to the
classical strategies where she hides her query among randomly chosen
ones, the security of the QPQ does not rest on the classical randomness
of the queries. This is evident from the simplest version of the
protocol, where the single query $j$ is answered.  However, this
classical randomness is a useful resource also for QPQ, since Alice can
increase the probability of catching a cheating Bob by choosing a high
number of random queries in her superposition.

The user security of the protocol rests on two key features, namely, the
fact that Alice is sending her queries in random order, and the fact
that she is sending them one by one. The first feature prevents Bob
from knowing which kind of query he is receiving at each time: if he
knew when the superposed queries are arriving, he would just let them
through without measuring them and measure the other queries, finding
out $j$ and evading detection. The second feature prevents Bob from
employing joint measurements on the queries. In fact, if he was
allowed joint measurements, he would find out the value of $j$ since
the subspaces spanned by the joint states of Alice's queries are
orthogonal for different choices of $j$.  

To discuss the user security of the protocol it is worth starting from a
simple cheating strategy.  Suppose for instance that Bob performs
projective measurements on both of Alice's queries. By doing so he
will always recover the value of $j$. Moreover with probability $1/2$,
one of his two measurement results will return $0$ in correspondence
to Alice's superposed query.  In this case, Bob's attempt at cheating
is successful, as he can correctly re-prepare both of Alice's queries.
However, with probability $1/2$, Bob gets $j$ from both measurements,
and it will impossible for him to determine which was the order of
Alice's queries.  In this case, no strategy of his has more than $1/2$
probability of passing Alice's test. In fact, this is the probability
that a state of the form $|j\rangle_Q|A_j\rangle_R$ passes the test of
being of the form $(|j\rangle_Q|A_j\rangle_R
+|0\rangle_Q|0\rangle_R)/\sqrt{2}$. If Bob uses this cheating
strategy, Alice can find it out with probability $1/4$ (this number
can be easily increased using the modified QPQ protocols discussed
above).

What if Bob employs a more sophisticated cheating strategy? Bob is
presented randomly with one among two possible scenarios ($A$ or $B$)
depending on which state Alice sends first. These scenarios refer to
the following joint states of her query
$|S_A\rangle=|j\rangle_{Q_1}(|j\rangle_{Q_2} +
|r\rangle_{Q_2})/\sqrt{2}$ and $|S_B\rangle=(|j\rangle_{Q_1} +
|r\rangle_{Q_1})|j\rangle_{Q_2}/\sqrt{2}$, where $Q_1$ and $Q_2$ are
her first and second query. The failure of the above cheating strategy
stems from Bob's impossibility to determine which scenario Alice is
using. This is a common problem to all cheating strategies: it is
related to the non-orthogonality of the states $|S_A\rangle$ and
$|S_B\rangle$, and to the limit posed by the timing of the protocol
(to gain access to $Q_2$, Bob must first respond to $Q_1$).  Working
along these lines, one can show that Alice has a nonzero probability
of discovering that Bob is cheating, whatever sophisticated methods he
employs.  More precisely, following a derivation which is similar to that
performed in Ref.~\cite{CWINTER},
it can be shown that his impossibility of
performing joint measurements on $Q_1$ and $Q_2$ places a bound on the
information Bob obtains on $j$: Alice can enforce the privacy of her
queries by requiring that Bob is never caught cheating. Here we just
sketch the main idea of the security proof, providing the details
elsewhere.

Any action by Bob in response to Alice's two queries can be described
in terms of two unitary transformations $U_1$ and $U_2$. The
transformation $U_1$ acts on the registers $Q_1$, $R_1$ and on an
ancillary system $B$ which is under Bob's control (it also includes
his database). The transformation $U_2$ acts on $Q_2$, $R_2$ and $B$.
If Bob is not cheating, $U_1$ and $U_2$ are instances of the qRAM
algorithm of Eq.~(\ref{inout}) below: they coherently copy the
information from the database to the $R$ registers leaving the ancilla
$B$ in its initial state. If instead Bob is cheating, at the end of
the communication the system $B$ will be correlated with the rest.  In
this case Alice's final state is the mixture
\begin{eqnarray}
\rho_{\ell}(j)\equiv\mbox{Tr}_B
\big[U_2 U_1 
|\Psi_{\ell}(j)\rangle\langle\Psi_{\ell}(j)|U_1^\dag U_2^\dag\big],
\end{eqnarray}
where the label $\ell=A,B$ refers to the scenario used by Alice to
submit her query $j$, and where
$|\Psi_\ell(j)\rangle\equiv|S_\ell\rangle_{Q_1Q_2}|0\rangle_{RB}$ is
the corresponding input state ($|0\rangle_{RB}$ being the initial
state of the registers $R_{1,2}$ and of the ancilla $B$). The
probability $1-P_\ell(j)$ that the state $\rho_\ell(j)$ supplied by
Bob will pass Alice's test can be easily computed by considering its
overlap with the states corresponding to the answer that a
non-cheating Bob would provide. On Bob's side, the information $I_B$
that he retains on the query is stored in the final state of the
ancilla $B$, i.e.
\begin{eqnarray}
\sigma_\ell(j)\equiv\mbox{Tr}_{Q_1Q_2R_1R_2}\left [U_2 U_1
  |\Psi_\ell(j)\rangle\langle\Psi_\ell(j)|U_1^\dag U_2^\dag\right]
\labell{bst}\;.
\end{eqnarray}
An information-disturbance trade-off~\cite{chuang} can be obtained by
noticing that if $1-P_{\ell}(j)\simeq 1$, then $\sigma_{\ell}(j)$ must
be independent from $j$. Specifically, requiring
$P_{\ell}(j)\leqslant\epsilon$ for all $\ell$ and $j$, one can show
that $1-F(\sigma_\ell(j),\sigma_*) \leqslant O(\epsilon^{1/4})$, where
$\sigma_*$ is a fixed state and $F$ the fidelity~\cite{fidelity}.
Therefore, in the limit of $P_{\ell}(j)\rightarrow 0$ (i.e.~Bob passes
the test with high probability), we see that the states he retains are
independent from the label $j$. This can also transformed into an
upper bound on the mutual information $I_B$ evaluating the Holevo
information~\cite{holevo} associated to the ensemble $\{ p_j
,\sigma(j)\}$ where $p_j=1/N$ is the probability that Alice will send
the $j$-th query, and where $\sigma(j)=[\sigma_A(j)+\sigma_B(j)]/2$ is
the final state of $B$ (from his point of view), since Alice randomly
chooses among the scenarios $A$ and $B$ with probability $1/2$.  By
doing so it can be shown~\cite{security} that $I_B\leqslant
O(\epsilon^{1/4}\;\log_2N)$.

In closing, we comment on the quantum random access memory (qRAM)
algorithm~\cite{qram,chuang} that Bob uses to interrogate his database
while preserving coherence, as required by the QPQ protocol. The aim
of the qRAM protocol is to read, in a memory array, a location
specified by an index register $Q$, and return the contents in a
second register $R$. The register $Q$ may contain a quantum
superposition of location addresses. The content of the $n$-qubit
address-register $Q$ is correlated by a unitary transformation $U$ to
the spatial position of a single qubit, which acts as a data bus. This
means that the binary encoding in the quantum register is translated
into a unary encoding on the location of the bus qubit, which is thus
into one of $2^n$ possible locations (or in more than one location in
quantum superposition). Now the qubit locally interacts with the
memory cell array, and the addressing procedure is reversed by running
the binary-to-unary encoding $U$ protocol backwards (an
``uncomputation'' performed by the unitary $U^\dag$).  This
decorrelates the position of the bus qubit from the $Q$ register
(otherwise quantum coherence would be destroyed).  Its internal state
contains the value of the memory cell (cells) that was to be read.
Essentially, the qRAM algorithm implements the transformation
\begin{eqnarray}
\sum_{j} \alpha_j |j\rangle_Q \rightarrow \sum_{j} \alpha_j
|j\rangle_Q |A_j\rangle_R\labell{inout}\;,
\end{eqnarray}
where $A_j$ is the content of the $j$th memory location, and
$\alpha_j$ are arbitrary amplitudes.

Conventional designs for quantum random access memory based on
classical architectures~\cite{chuang} require $O(2^n)$ quantum logic
operations to perform a qRAM call.  However, we have recently
exhibited qRAM designs in which the number of quantum logic operations
to perform a call can be reduced to $O(n)$~\cite{qram}. Hence,
constructing a qRAM for quantum private queries should be
significantly easier than constructing a large-scale quantum computer.

\acknowledgments V. G. acknowledges the support of Quantum Information
Research program of the Centro di Ricerca Ennio De Giorgi of Scuola
Normale Superiore.  L. M. acknowledges support from MIUR through PRIN
2005 and from EU through the project SECOQC (IST-2003-506813).


\begin{thebibliography}{99}
\bibitem{NOTA}Slightly better performances can be obtained by assuming
  the existence of multiple non-mutually communicating replicas of the
  servers, see Refs.~\cite{PIR}. Moreover sub-linear communication
  complexity can be achieved under the some computational complexity
  assumption --- see for instance E. Kushilevitz, R. Ostrovsky in
  Proceedings of Thirty-eigth Annual IEEE Symposium on the Foundations
  of Computer Science (FOCS-97).
\end{thebibliography}
\end{document}